\documentclass{aa}
\usepackage{epsfig}
\begin{document}

%
   \title{The ARGO-YBJ Detector and high energy GRBs}


\author {
C.~Bacci\inst{1}, K.Z.~Bao\inst{2},  F.~Barone\inst{3}, B.~Bartoli\inst{3}, 
D.~Bastieri\inst{4}, P.~Bernardini\inst{5}, R.~Buonomo\inst{3},
S.~Bussino\inst{1},
E.~Calloni\inst{3}, B.Y.~Cao\inst{6}, R.~Cardarelli\inst{7}, 
S.~Catalanotti\inst{3}, A.~Cavaliere\inst{7}, F.~Cesaroni\inst{5}, 
P.~Creti\inst{5}, Danzengluobu\inst{8}, B.~D'Ettorre Piazzoli\inst{3}, 
M.~De Vincenzi\inst{1}, T.~Di Girolamo\inst{3}, G.~Di Sciascio\inst{3}, 
Z.Y.~Feng\inst{9}, Y.~Fu\inst{6}, X.Y.~Gao\inst{10}, Q.X.~Geng\inst{10}, 
H.W.~Guo,\inst{8}, H.H.~He\inst{11}, M.~He\inst{6}, Q.~Huang\inst{9},
M.~Iacovacci\inst{3}, N.~Iucci\inst{1}, H.Y.~Jai\inst{9},
F.M.~Kong\inst{6}, H.H.~Kuang\inst{11}, Labaciren\inst{8}, B.~Li\inst{2}, 
J.Y.~Li\inst{6}, Z.Q.~Liu\inst{10}, H.~Lu\inst{11}, X.H.~Ma\inst{11}, 
G.~Mancarella\inst{5}, S.M.~Mari\inst{12}, 
G.~Marsella\inst{5}, D.~Martello\inst{5}, D.M Mei\inst{8}, X.R.~Meng\inst{8},
L.~Milano\inst{3}, A.~Morselli\inst{7}, J.~Mu\inst{10}, M.~Oliviero\inst{13},
P.~Padovani\inst{7}, M.~Panareo\inst{5}, M.~Parisi\inst{1}, 
G.~Pellizzoni\inst{1}, Z.R.~Peng\inst{11}, 
P.~Pistilli\inst{1}, R.~Santonico\inst{7}, 
G.~Sartori\inst{4}, C.~Sbarra\inst{4}, G.~Severino\inst{13}, 
P.R.~Shen\inst{11}, R.~Sparvoli\inst{7}, 
C.~Stanescu\inst{1}, J.~Su\inst{11}, L.R.~Sun\inst{2}, S.C.~Sun\inst{2}, 
A.~Surdo\inst{5}, Y.H.~Tan\inst{11}, S.~Vernetto\inst{14}, M.~Vietri\inst{1}, 
C.R.~Wang\inst{6}, H.~Wang\inst{11}, H.Y.~Wang\inst{11}, Y.N.~Wei\inst{2}, 
H.T.~Yang\inst{12}, Q.K.~Yao\inst{2}, G.C.~Yu\inst{9}, X.D.~Yue\inst{2}, 
A.F.~Yuan\inst{8}, H.M.~Zhang\inst{11}, J.L.~Zhang\inst{11}, 
N.J.~Zhang\inst{6}, T.J.~Zhang\inst{10}, X.Y.~Zhang\inst{6}, 
Zhaxisangzhu\inst{8}, Zhaxiciren\inst{8}, Q.Q.~Zhu\inst{11}}

\institute{
INFN and Dipartimento di Fisica dell'Universit\`a di Roma Tre, Italy\
\and
Zhenghou University, Henan, China\
\and
INFN and Dipartimento di Fisica dell'Universit\`a di Napoli, Italy\
\and
INFN and Dipartimento di Fisica dell'Universit\`a di Padova, Italy\
\and
INFN and Dipartimento di Fisica dell'Universit\`a di Lecce, Italy\
\and
Shangdong University, Jinan, China\
\and
INFN and Dipartimento di Fisica dell'Universit\`a di Roma "Tor Vergata", Italy\
\and
Tibet University, Lhasa, China\
\and
South West Jiaotong University, Chengdu, China\
\and
Yunnan University, Kunming, China
\and
IHEP, Beijing, China\
\and
Universit\'a della Basilicata, Potenza, Italy\
\and
Osservatorio Astronomico di Capodimonte, Napoli, Italy\
\and
Istituto di Cosmogeofisica del CNR and INFN, Torino, Italy\\
              email: vernetto@lngs.infn.it}

   \offprints{S. Vernetto}

   \date{Received December 15, 1998; accepted 22 April 1999}

\authorrunning{C.Bacci et al}
\titlerunning{The Argo-YBJ Detector and high energy GRBs}

   \maketitle

   \begin{abstract}

ARGO-YBJ (Astrophysical Radiation with Ground-based Observatory at YangBaJing) 
is a detector optimized to study small size air showers.
It consists of a layer of Resistive Plate Counters (RPCs)
covering an area of $\sim$ 6500 m$^2$
and will be located in the Yangbajing Laboratory (Tibet, China)
at 4300 m a.s.l. .
ARGO-YBJ will be devoted to a wide range of fundamental issues in cosmic 
rays and  astroparticle physics,
including in particular gamma-ray astronomy and gamma-ray bursts physics 
in the range 10~GeV$\div$500~TeV.
The sensitivity of ARGO-YBJ to detect high energy GRBs is 
presented. 

   \end{abstract}

\section{Introduction}

The study of the GeV-TeV component of gamma-ray bursts is of great
importance to understand the acceleration mechanisms and the
sources physical conditions.
The detection of GeV gamma-rays by EGRET during some intense GRBs \cite{egret}
suggests the possibility that a high energy component could be present
in all events. Furthermore several models predict GeV and TeV emission,
sometimes correlated with UHECRs production (see \cite{baring}, for a review).
Due to the low fluxes and
the small sensitive areas of satellite experiments,
gamma-rays of energy larger than a few tens of GeV, 
must be detected by ground based experiments located 
at mountain altitude measuring the secondary particles generated by 
gamma-rays in the atmosphere.
At energies E~$<$~10~TeV the number of particles reaching the ground  
is to small to reconstruct the shower parameters using 
a standard air shower array, made of several detectors spread over 
large areas. 
On the contrary, a detector consisting of a full coverage layer of counters, 
providing a high granularity sampling of all particle showers, can
succesfully measure arrival direction and primary energy of 
small showers, allowing the study of the unexplored range of gamma energies 
between 20 GeV and 300 GeV \cite{proposal}. 

\section {The ARGO-YBJ detector}

ARGO-YBJ is an air shower detector optimized to observe small size showers,
to be constructed in the Yangbajing Laboratory 
(Tibet, China) at an altitude of 4300 m a.s.l.
The experiment consists of a $\sim$ 71$\times$74 m$^2$ core detector
realised with a single layer of RPC's ($\sim 90\%$ of active 
area), surrounded by an outer detector ($\sim 30\%$ of active area)  
for a total size of $\sim$ 100$\times$100 m$^2$.
A lead converter 0.5 cm thick will cover uniformly the RPC plane in order
to increase the number of charged particles by conversion of shower photons
in $e^{\pm}$ and to reduce the time spread of the shower front \cite{addendum}.
ARGO-YBJ can image with high efficiency and sensitivity atmospheric showers 
initiated by primaries of energies in the range 10~GeV$\div$~500~TeV.
Its main physics goals are \cite{proposal}: Gamma-astronomy at 
$\sim$~100~GeV energy threshold, with a sensitivity to detect unidentified 
point sources of intensity as low as $10\%$ of the Crab Nebula;  
Gamma-Ray Burst physics, extending the satellite measurements
at energies E $>$ 10 GeV; $\overline{p}/p$ ratio in the TeV energy range;
Sun and heliosphere physics.
The detector assembling should start in 2000 and data taking with the
first $\sim 750$ $m^2$ of RPC's in 2001.

\section {Sensitivity to high energy GRBs}

A high energy GRB is detectable if the number
of air showers from the gamma-rays is significant larger than
the fluctuations of the background, due to showers from cosmic rays with 
arrival directions compatible with the burst position. 
A good angular resolution is of major importance in order to reduce 
the background and increase the detection sensitivity.
The  angular resolution  and the effective area of ARGO-YBJ to detect 
gamma-rays as a function of the energy have been obtained by 
means of simulations. 
For gamma-rays with energy as low as E $\sim$ 10-20 GeV,
the opening angle around the
source direction in which 70$\%$ of the signal showers are contained
is $\sim$ 5$^{\circ}$. 

To evaluate the sensitivity of ARGO-YBJ to detect GRBs, we considered
a burst with a power law energy spectrum $dN/dE \propto E^{-\alpha}$ 
extending in the range 1~GeV~$\div E_{max}$,
a duration $\Delta t$=1 s, and a zenith angle $\theta$=20$^{\circ}$.
The burst will give a signal with a significance 
larger than 4 standard deviations if the energy fluence in 
the range 1~GeV~$\div E_{max}$ is larger than a minimum value 
F$_{min}$. Fig.1 shows F$_{min}$ as a function of $E_{max}$
for 3 spectral slopes. For a generic duration $\Delta t$
the minimum fluences detectable are given by $F_{min} \sqrt{\Delta t}$.

   \begin{figure}
\vspace{0cm}
\hspace{0cm}\epsfxsize=8.8cm \epsfbox{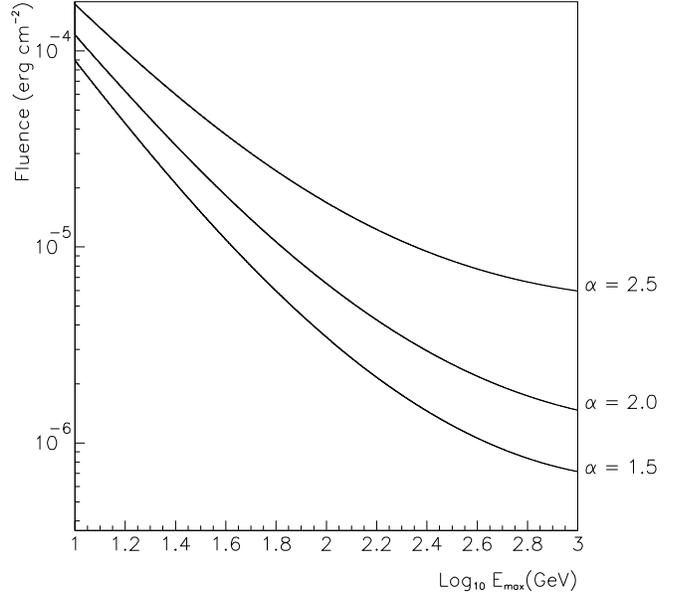}
\vspace{0cm}
      \caption[]
{The minimum energy fluence in the
range 1~GeV$\div E_{max}$ 
of a GRB detectable by ARGO-YBJ as a function of the
maximum energy of the spectrum $E_{max}$ for 3 spectral slopes.}
         \label{gauss}
   \end{figure}

In the energy range considered the sensitivity is strongly dependent 
on the maximum energy of the spectrum $E_{max}$.
ARGO-YBJ can observe GRBs with energy fluences of a few 10$^{-6}$ 
erg cm$^{-2}$ if the energy spectrum extends at least 
up to $\sim$~200~GeV with a slope $\alpha \leq$2;
the minimum detectable fluence is $\sim$ 10$^{-5}$ if $E_{max}~\sim$~30~GeV.

This is of particular importance, since if GRB sources are located at 
cosmological distances, the high energy tail of the spectrum is affected
by the $\gamma \gamma \rightarrow e^+ e^-$ interaction of gamma-rays
with low energy starlight photons in the intergalactic space.
According to \cite{abso}, at a distance corresponding to a redshift
$z$=0.1 the absorption is
almost negligible, while at $z$=0.5 (1.0) the absorption becomes 
important for photons of energy E~$>$~100~(50)~GeV.
These values give an idea of the possible maximum energy of the
GRBs spectra as a function of their distance, and from Fig.1 one can infer
the maximum sensitivity of ARGO-YBJ to detect cosmological GRBs.
The minimum observable fluences can be compared with the fluences
measured by EGRET in the 1 MeV-1 GeV energy range: $F \sim
10^{-5} \div 10^{-4}$ erg cm$^{-2}$ \cite{egret}. 
Since EGRET spectral slopes $\alpha$
are mostly  $\sim$ 2, one could expect fluences of 
the same order of magnitude at energies above 1 GeV. 
From Fig.1 one can conclude that ARGO-YBJ could detect GRBs 
with the same intensity of those observed
by EGRET provided that the energy spectrum extends up to few tens of GeVs;
the sensitivity increases by a factor $\sim$10 for
spectra extending up to $E_{max} \sim$~200~GeV.


\begin{thebibliography}{}

\bibitem[(Abbrescia et al., 1996)]{proposal}
Abbrescia et al., Proposal of the ARGO experiment, 1996,
http://www1.na.infn.it/wsubnucl/cosm/argo/argo.html
\bibitem[(Bacci et al., 1998)]{addendum}
C. Bacci et al., Addendum to the ARGO Proposal, 1998,
http://www1.na.infn.it/wsubnucl/cosm/argo/argo.html
\bibitem[Baring 1997]{baring}
Baring M.W., astro-ph9711256, 1997
\bibitem[(Catelli et al., 1997)]{egret}
Catelli J.R. et al., AIP Conf.Proc.428, 309, 1997
\bibitem[Salomon and Stecker (1998)]{abso}
Salomon M.H. and Stecker F.W., 1998, ApJ 493, 547
\end{thebibliography}
\end{document}